\documentclass[orivec]{llncs}

\usepackage{xspace}
\usepackage{tabularx}
\usepackage{multicol}
\usepackage{xcolor}
\usepackage{paralist}
\usepackage{wrapfig}
\usepackage[utf8]{inputenc}
\usepackage[T1]{fontenc}
\usepackage{color}
\usepackage[inline]{enumitem}

\usepackage{amsmath,amssymb}

\usepackage{tikz}
\usetikzlibrary{shapes,arrows}

\usepackage[hide]{ed}

\usepackage{basics}

\usepackage[bookmarks,linkcolor=blue,citecolor=blue,urlcolor=gray,colorlinks,breaklinks,bookmarksopen,bookmarksnumbered]{hyperref}

\begin{document}

\title{A Survey of Languages\\ for Formalizing Mathematics}
\author{
Cezary Kaliszyk\thanks{Supported by ERC starting grant no.\ 714034 \textit{SMART}}\orcidID{0000-0002-8273-6059}
  and
Florian Rabe\thanks{Supported by DFG grant RA-1872/3-1 OAF and EU grant Horizon 2020 ERI 676541 OpenDreamKit.}\orcidID{0000-0003-3040-3655}}
\institute{Universities of Innsbruck resp. Erlangen-Nürnberg}

\maketitle

\begin{abstract}
  In order to work with mathematical content in computer systems, it is necessary to
  represent it in formal languages.
  Ideally, these are supported by tools that verify the correctness of the content, allow computing with it, and produce human-readable documents.
  These goals are challenging to combine and state-of-the-art tools typically have to make difficult compromises.\\[1mm]
  In this paper we discuss languages that have been created for this purpose, including
  logical languages of proof assistants and other formal systems, semi-formal languages,
  intermediate languages for exchanging mathematical knowledge, and language frameworks that allow building customized languages.\\[1mm]
  We evaluate their advantages based on our experience in designing and applying languages and tools for formalizing mathematics.
  We reach the conclusion that no existing language is truly good enough yet and derive ideas for possible future improvements.
\end{abstract}

\section{Introduction}\label{sec:intro}
Today's formal systems can verify advanced
theorems in mathematics~\cite{Hales-Developments}, such as the Kepler
conjecture~\cite{halesdense} or the Feit--Thompson theorem~\cite{DBLP:conf/itp/GonthierAABCGRMOBPRSTT13}, as well as certify important computer systems,
such as the CompCert C compiler~\cite{Compcert09} and
the seL4 microkernel~\cite{SeL414}.
All these systems and projects use advanced logical languages that are computer-understandable but hard for humans to write and read.

Computer science commonly defines and implements such \emph{formal} languages for mathematical content that define syntax and semantics and offer strong automation support.
However, non-trivial and expensive transformation steps are needed to formalize human-near \emph{natural} language texts in them.

This is in contrast to standard approaches to writing mathematics or specifying computer systems, which use \emph{natural} language with interspersed syntactically unrestricted formulas, e.g., as written in \LaTeX.
While interpreting this natural language is very difficult for computers (arguably AI-complete), it is extremely effective for humans in a way that formal languages have so far not been able to capture.
In fact,  in 2007, Wiedijk claimed~\cite{qedrevisited}, citing four representative statements, that no existing formal system was sufficient to naturally express basic mathematical content.
Despite the progress made since then, his critique still applies.

We give an introduction to the objectives and main approaches in Section~\ref{sec:overview}.
Then Sections \ref{sec:fola} and \ref{sec:infola} describe the main approaches: formal system and intermediate languages in more detail. Sections~\ref{sec:languageframeworks} and \ref{sec:lib} describe closely related orthogonal aspects: language frameworks and interchange libraries. We evaluate our findings and conclude in Section~\ref{sec:conc}.

\section{Overview}\label{sec:overview}
\subsection{Objectives}

Thus, a big picture goal of the field is a tighter integration of (i) natural language mathematical content such as textbooks or software specifications, and (ii) formalization of such content in logics and theorem proving systems.
We can identify the following overarching objectives:

\paragraph{A universal formal language for mathematical content that supports complex structuring mechanisms}
We want a language that combines the universality of natural mathematical languages with the automation support of formal logics and programming languages.
It should be closer to mathematics than these formal languages in regards to abstract syntax, notations, and type system.
This is critical not only for generality but also to appeal to mathematicians at all because, as Wiedijk observes, most mathematicians do not like to read (or write) code~\cite{qedrevisited}.
On the other hand, it should be fully formal including automation support for type, module, and proof systems of formal languages that have proved critical for large scale applications.

\paragraph{A comprehensive standard library of mathematical concepts}
The language must allow for building a standard library of mathematical concepts.
In order to allow for semantics-aware machine support, it should be more formal than existing informal libraries such as induced by Wikipedia or PlanetMath by including formal types, notations, and properties.
On the other hand, in order to achieve generality and support interoperability, it should not be committed to a particular logic like all the major formal libraries are.
This combination of advantages would allow it to serve as a \emph{standard} library, i.e., a central community resource to be used, e.g., to cross-link between existing libraries in a star-shaped network or to provide a basis for projects like FAbstracts~\cite{fabstracts}.

\paragraph{Practical workflows that integrate natural and formal languages}
Such a language and standard library would enable substantially better tool support for working researchers in mathematical sciences:
Being structurally similar to both natural and formal languages, they could serve as an interface language for tools of either kind.
This would allow enriching existing workflows such as \latex-based authoring or proof-assistant--based verification.
For example, researchers could easily formalize conjectures and their proof outlines in a general language as a first and cheap formalization step, before or instead of a full verification of the proof.
This would avoid hindering practicians as today's all-or-nothing approach of formalization in proof assistants tends to do~\cite{KR:qed:14}.
It is critical for success here to retain existing workflows instead of trying to develop a single be-all-end-all tool that no one would adopt.
Therefore, any major project in this direction must aim at developing concrete improvements to the current ecosystem.

\subsection{Approaches}

The most successful formal languages for mathematical content have been developed in the areas of formal logic where they occur most prominently as the input languages of proof assistants as well as in computer algebra where they occur as programming languages fitted to mathematical algorithms.
These combine formal foundations with complex structuring mechanisms, especially type, module, proof, and computation systems, which have proved critical to achieve efficient large scale tool support.
Importantly, these fix not only the syntax but also the semantics of, e.g., proofs and computations
On the contrary, in natural language, these are not spelled out at all, let alone explicated as a primitive feature --- instead, they are emergent features driven by conventions and flexibly adaptable to different contexts.
Consequently, formalization is usually a non-structure-preserving transformation that is often prohibitively expensive and creates an entry barrier for casual users.
For example, the mathematician Kevin Buzzard admonishes computer scientists to build more human-near languages ``so users can at least read sentences they understand and try to learn to write these sentences''.

Between the extremes of natural and formal languages, a variety of \emph{intermediate} languages make different trade-offs aiming at combining the universal applicability of natural language with the advantages of formal semantics.
A central observation is that
\begin{itemize}
 \item existing intermediate languages apply only the formal syntax and (to varying degrees) semantics of formal languages but not their complex structuring mechanisms, and
 \item this limitation is not necessarily an inherent feature of the approach but rather a frontier of research.
 \end{itemize}
 The following table summarizes the resulting trichotomy and shows how each kind of language satisfies only two out of three essential requirements:

\begin{center}
\begin{tabular}{l|ccc}
language properties      & natural & intermediate & formal \\
\hline
formal syntax and semantics & -- & + & + \\
complex structuring  & + & -- & + \\
universal applicability  & + & + & -- \\
\end{tabular}
\end{center}

In the subsequent sections, we discuss the state of the art for these languages in more detail.

\section{Formal Languages}\label{sec:fola}

Formal languages use a wide variety of foundations and complex structuring mechanisms.
This unfortunately means that it is rare for two tools to be compatible with each other.
Additionally, all are quite removed from natural language.
In the sequel, we discuss the most important complex structuring features.

\subsection{Type Systems}

Many formal systems use what we call \emph{hard} type systems, which assign a unique type to each object and are thus easiest to automate.
Systems derived from Martin-L\"of type theory \cite{martinlof} or the calculus of constructions \cite{calcconstructions}
usually use the proofs-as-programs correspondence (Curry-Howard \cite{curry,howard}) that represents mathematical properties as types and proofs as data.
These include Agda \cite{agda}, Coq \cite{coq}, Lean \cite{lean}, Matita \cite{matita} as well as Nuprl \cite{nuprl}.
Systems derived from Church's higher-order logic \cite{churchtypes} usually use the LCF architecture \cite{lcf} that uses an abstract type of proved theorems.
These include HOL4 \cite{hol4}, ProofPower \cite{proofpower}, Isabelle \cite{isabellehol}, and HOL Light \cite{hollight}.

Hard type systems are at odds with natural language as the unique-type property precludes representing mathematical sets and subsets as types and subtypes.
In particular, the lack of expressive subtyping in hard type systems is fundamentally at odds with every day mathematics, where sets and subsets are used throughout: hard type system precludes a direct representation of sets as types because they cannot represent the rich (even undecidable) subset relation using subtyping.

Multiple systems have explored compromises.
We speak of \emph{semi-soft} type systems if a hard type system is extended with variants of subtyping.
For example, PVS \cite{pvs} uses predicate subtypes, Lean \cite{lean} and Nurpl \cite{nuprl} support predicate subtypes and quotient types, and IMPS uses \cite{imps} refinement types.

Both hard and semi-soft type systems force users to choose between representing information using the type system (e.g., $\forall x:\N.P(x)$) or the logical system (e.g., $\forall x. x\in \N \Rightarrow P(x)$).
Problematically, this choice usually has far-reaching consequences, e.g., the type system may be decidable but the logic system undecidable.
But from the perspective of mathematics this distinction is artificial, and the fact that the two resulting representations may be entirely incompatible down the road is very awkward.

These problems are avoided in \emph{untyped} languages.
ACL2 \cite{acl2} is a first-order logic on top of the untyped $\lambda$-calculus of
Lisp that strongly emphasizes computation.
Untyped set theory is used in Isabelle/ZF \cite{isabelle_zf}, Metamath \cite{metamath}, and the B method \cite{bmethod}.
Untyped languages are also common in virtually all computer algebra systems, such as Mathematica \cite{mathematica} or SageMath \cite{sage}.

We speak of \emph{soft type systems} if unary predicates on the untyped objects mimic types and the type system is an emergent feature of the logical system.
These are used most prominently in Mizar \cite{mizar} among proof assistants and GAP \cite{gap} among computer algebra systems.
Both use the types-as-predicates approach, where the semantics of a type is given by a unary predicate ranging over untyped objects.
Both allow declaring functions and dependent types (i.e., predicates $n+1$ arguments that return a unary predicate after fixing the first $n$ arguments) that have type constraints on the arguments.
Soft type systems are generally hardest to automate because type-checking is reduced to undecidable theorem proving.
Here Mizar leverages theorem proving: the type checker is guided by user-stated typing rules (called registrations), which are specially marked theorems about typing properties.
GAP leverages computation: the typing predicates must be computable properties, which are computed and cached at run-time for every object.

Thus, soft type systems are heuristic, which makes implementations more difficult for the developer and their behavior less predictable for the user.
But they are the most human-friendly.
A combination of hard and soft type systems, where advanced hard type systems are emergent features built systematically on top of a soft one, could potentially model mathematical content best but has so far received much less systematic attention than the above approaches.
But as theorem proving technology becomes more routine, they become more and more attractive.

An example soft type system has been recently developed on top of a hard-typed Isabelle for the Isabelle/Mizar object logic~\cite{ckkp-jar19},
which expresses the largest softly typed proof library in a logical framework.
As part of this research, the type system of Mizar has been formalized including
its intersection type constructions, 
various ways to express
set-theoretic structures, 
and declarative proof
translations~\cite{ckkp-itp19} have been investigated. Furthermore, a common
foundation for proofs that allows practically combining results between HOL and
set theory has been developed~\cite{cbckkp-itp19}.

\subsection{Module Systems}

A key use of modules is to represent structures (also called records or theories); here the abstract definition (e.g., ``group'') is represented as a module, and concrete models (e.g., individual groups) are represented as instances of the module.
The used module systems vary widely but roughly fall into the following groups.

Firstly, ML-inspired \emph{external} module systems use a two-layer language where the module language is external to the logical language.
These allow for inheritance, refinement, and instantiation (e.g., Coq modules, Isabelle locales, PVS theories, SageMath categories, Axiom categories) as well as more advanced structuring such as parametric modules (e.g., PVS), module expressions (e.g., Isabelle locales, Axiom joins), or morphisms between modules (Isabelle, IMPS).
Hard module systems differ greatly from natural language where no two-layer language is fixed.

Secondly, \emph{internal} module systems use record types to mimic modular structure inside the type system.
This is possible in all systems that support record types (e.g., Agda, Coq, Isabelle, Lean, PVS); Mizar's structures behave similarly.
Soft modules are more flexible and thus similar to natural language, but the lack of a concise module system makes modular reasoning like inheritance and refinement more difficult.
For example, soft module systems must manually employ extra-logical conventions (e.g., \cite{gonthier_packaging}), and combining modules built with different conventions quickly becomes impractical. This is even worse in the common case where both hard and soft module systems are present in parallel (we have initiated work in this direction in \cite{MRK:modelsof:18}).

Both of the above can be seen as \emph{hard} module systems in the sense that a module encapsulates a fixed set of declarations that induce selectors that can be applied to the module's instances.
A third group, which we call \emph{soft} module systems is somewhat hypothetical as it is used much less widely.
Here, in analogy to soft tying, modules are treated as unary predicates that range over objects.
Inheritance then becomes a special case of implication.
This idea is used in the GAP module system, whose soft types (called properties) and soft modules (called categories) are treated very similarly: they are jointly filters, and the run-time system tracks which object satisfies which filters.
The main difference between them is that categories can have constructors and thus allow for filters that are satisfied by construction.

Finally, since module systems have mostly been designed as extensions of existing logical languages, both hard and soft module systems fail to capture a number of essential features of natural mathematical language:
the identification of isomorphic instances of the same module;
the seamless extension of operations across substructures and quotient structures (e.g., $+$ is first defined on $\N$, then extended to $\Z$);
the flexibility of presence and order of fields in a structure (e.g., $(\Z,+,*)$ and $(\Z,+,0,-,*,1)$ should be the same ring);
the context-sensitive meaning of structures (e.g., $\Z$ should be a ring or a total order, depending on the context);
and in many systems also the implicit application of forgetful functors (e.g., a group is not automatically also a monoid).

\subsection{Proof Systems}

Formal languages shine when using logics implemented in proof assistants to find and check proofs automatically.
\emph{Tactic-based} proof systems (e.g., HOL Light) are optimized for efficiency of proof checking but have an imperative flavor that is very different from natural mathematical language.
\emph{Declarative} proof systems (e.g. Mizar, Isabelle/Isar) were designed to be closer to natural language.

While many current tools support declarative proofs using quite similar languages, all of these are intertwined with the respective logic and therefore not immediately reusable as a universally applicable declarative proof language.
In particular, the expressivity of these languages is limited by the strength of the underlying logic, i.e., they can only express the kind of proof steps that can be potentially verified by the tool.
Declarative proof languages are conceptually close to natural mathematics but technically tied to specific logics.
We discuss logical languages in more detail in section~\ref{sec:languageframeworks}.

In computer algebra systems no formal logics are implied and automated reasoning is restricted to computable properties.
Additionally, these systems can capture logical properties by user declaration:
for example, most systems' libraries distinguish between groups and commutative groups and allow users to construct a group as commutative even if that property is not proved.

\subsection{Computation Systems}
The second major application of formal languages stems from computer algebra systems, which use mathematics-customized variants of general purpose programming languages for efficient computation.

Even though mathematics uses mostly pure functions, most systems are based on Turing-complete imperative programming, mostly to reuse existing user knowledge and fast implementations.
It is common to use the same language for pure mathematical algorithms and interspersed imperative meta-operations like I/O, logging, memoization, or calling external tools (in particular in SageMath).

Proof assistants take a much more restricted approach to integrate pure computations with a logic.
Three main approaches exist.
Firstly, normalization in the type theory, in particular $\beta$-reduction is a primitive form of computation.
It becomes much stronger when combined with (co)inductive types and recursion, and these are primitive features in most complex type theories like Coq.
Systems then usually include heuristic termination criteria to check the soundness of the functions, which leads to a trade-off between logical and Turing-completeness.
Secondly, certain theorems such as Horn formulas about equality can be interpreted as conditional rewrite rules.
Typically, systems require the user to choose which theorems to use and then exhaustively rewrite expressions with them.
This is much slower but allows for a much simpler language as computation is relegated to the meta-level.
This is the main method used in systems without primitive (co)inductive types such as Isabelle.
Thirdly, computation can be supplied by external tools or special kernel modules.
This computation can be a part of the, consequently rather big, trusted code base, such as in PVS decision procedures, the usage of SAT solvers is Mizar~\cite{Naumowicz14}. This is also the case in Theorema: As the proof assistant is written in the Mathematica computer algebra system, it is in principle possible to use most Mathematica's algorithms inside Theorema~\cite{theorema}.
In some cases, a trade-off is possible where computations are run externally and their results are efficiently verified by the prover.

\section{Intermediate Languages}\label{sec:infola}

\usetikzlibrary{decorations,decorations.markings,decorations.text}
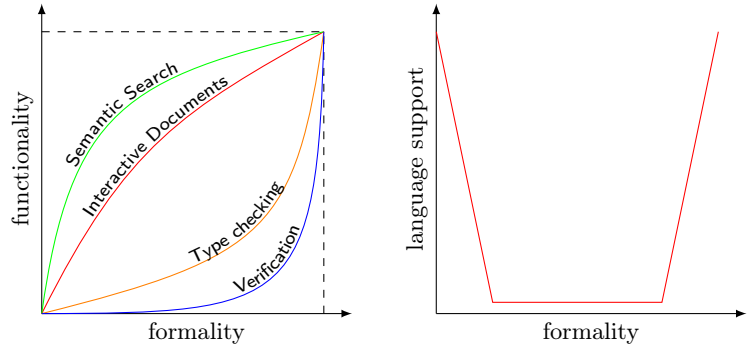
\begin{figure}[hbt]
\centering
\begin{tikzpicture}[scale=.75]
   \draw[|->, -latex] (0,0) -- coordinate (x axis mid) (5.5,0);
   \node[below=0.0cm] at (x axis mid) {formality};
   \draw[|->, -latex] (0,0) -- coordinate (y axis mid) (0,5.5);
   \node[rotate=90, above=0.0cm] at (y axis mid) {functionality};
  \draw[dashed] (0,5) -- (5,5);
  \draw[dashed] (5,0) -- (5,5);
  \draw[red,postaction={decorate,decoration={raise=0.3ex,text along path,
  text align=center,text={|\scriptsize\sffamily|Interactive Documents}}}]
(0,0) .. controls (1.5,3.0) and (2.5,3.6) .. (5,5);
  \draw[green,postaction={decorate,decoration={raise=0.3ex,text along path,
  text align=center,text={|\scriptsize\sffamily|Semantic Search}}}] (0,0) .. controls (0.5,3.6) and (1.5,4.2) ..
(5,5);
  \draw[orange,postaction={decorate,decoration={raise=0.3ex,text along path,
  text align=center,text={|\scriptsize\sffamily|Type checking}}}] (0,0) .. controls (4.0,1) and (4.5,1.5) .. (5,5);
  \draw[blue,postaction={decorate,decoration={raise=0.3ex,text along path,
  text align=center,text={|\scriptsize\sffamily|Verification}}}] (0,0) .. controls (4.4,0.01) and (4.9,0.1) ..
(5,5);
\end{tikzpicture}
\tb
\begin{tikzpicture}[scale=.75]
   \draw[|->, -latex] (0,0) -- coordinate (x axis mid) (5.5,0);
   \node[below=0.0cm] at (x axis mid) {formality};
   \draw[|->, -latex] (0,0) -- coordinate (y axis mid) (0,5.5);
   \node[rotate=90, above=0.0cm] at (y axis mid) {language support};

  \draw[red] (0,5) -- (1,.2) -- (4,.2)  -- (5,5);
\end{tikzpicture}
\caption{Functionality for intermediate languages (left) \newline market gap for stepwise formalization support (right)}\label{fig:intermediate}
\end{figure}

Intermediate languages try to capture the advantages of natural languages in a formal language.
There is a rather diverse set of such approaches, which we describe in groups below.
However, we can identify some general effects that motivate the design of many intermediate languages.

Firstly, an intermediate language can already provide sufficient automation support for some tasks.
Thus, it can serve as a more natural and easier-to-use target language for (partial) formalization if the task at hand is supported.
For example, search, interactive documents, or dependency management can be realized well in some intermediate languages and even benefit from structural similarity to the human-near natural language formulation.
The main counter-examples are verification and computation, which requires a lot more formalization.
This is indicated in Figure~\ref{fig:intermediate} (left).

Secondly, an intermediate language can serve as an interface between human-near natural language and a verification- or computation-oriented formal language.
This enables stepwise formalization and thus a smoother transition from the informal to the formal realm.
It may also allow for a separation of concerns where a domain experts transform content from informal to intermediate in a first step and a formalization transforms from intermediate to formal in a second step.
The relative lack of highly successful approaches in this style is indicated in Figure~\ref{fig:intermediate} (right).

Thirdly, the intermediate representation is often not or only barely committed to a particular formal language (e.g., a particular type system, module system, proof system or computation system).
During stepwise formalization, this means that the first step only needs to be done once and can then be reused for different second steps targeting different formal languages.
Expanding on this, we see that an intermediate language can provide an interoperability layer between formal languages.
That can help with the notorious lack of interoperability between formal systems (see also Section~\ref{sec:lib}).

\subsection{Controlled Natural Language}

These approaches combine formal grammars for fragments of natural language with formal languages for formulas.
Their goal is to make the surface syntax of a formal language as close to traditional mathematics as possible while retaining a formal grammar that allows for automated parsing.
While the languages often look similar to some semi-formal languages discussed below, we classify them differently because they use a formal logic-near language in their kernel.

This method is applied in MathLang \cite{mathlang,mathlang_handbook}, MathNat \cite{mathnat}, in \cite{kofler_parser} within the FMathL project, and Naproche \cite{naproche}.
Mizar \cite{mizar} is a logical language whose surface syntax has been carefully designed to look like a small fragment of natural language and thus looks similar to controlled natural language systems without being one.

These systems vary in how the semantics of the language is defined and how much implementation support is provided.
Both MathLang and Naproche use, effectively, a soft type system on top of set theory to define the semantics.
MathLang allows for translating content into proof assistants (Isabelle, Coq) for users to finish and check proofs, and Naproche uses automated first-order theorem provers to discharge proof obligations automatically.

Contrary to the verification/computation-oriented formal languages, where large libraries of formal content (up to $\sim 10^5$ theorems in the biggest systems) are developed and shared in vibrant communities, none of the controlled natural language systems provides a substantial library.
This is a hen-egg problem since large libraries often result from the practical necessities caused by verification and computation.
A critical limiting factor of existing controlled natural languages is the lack of scalable automation support and large libraries.

\subsection{Semi-Formal Languages}

These approaches aim at combining unrestricted natural mathematical language and formal language in the same document.
Contrary to the controlled natural language approaches discussed above, the interpretation of the natural language parts remains AI-complete.

\textbf{Flexiformal} systems use informal and formal language as alternatives, i.e., content may be written informally or formally.
Thus, not all mathematical content is formalized, and all tool support must degrade gracefully when informal and thus uninterpretable content is encountered.
The sTeX system \cite{stex} is a \latex package for annotating informal mathematical texts with its formal meaning, which then allows for writing (parts of) formulas in formal logical syntax.
In addition to pdf, sTeX documents can be processed into \omdoc documents \cite{omdoc}, which makes them available for further machine processing.
sTeX provides no type system and only a very simple hard module system.
It has been used by the developer to write his introductory computer science lecture notes.

\textbf{Literate programming} \cite{literate_programming} and approaches inspired by it allow for natural and formal language to appear in parallel.
Here, content is described twice: the formal version defines the semantics, and the informal version provides documentation.
Contrary to the other approaches mentioned here, this is not a third language in between formal and natural, but a combination of the two.
Several formal systems provide mature support for writing content in literate programming style such as Agda, Isabelle (see \cite{isabelle_documentoriented}), and Axiom.

\textbf{Discourse representation languages} \cite{KampReyle} perform an analysis of the language used in written mathematics and design a fixed set of disambiguation conventions.
Ganesalingam~\cite{Ganesalingam} proposes a system of types that together with a parsing procedure and the set of disambiguation conventions could be used to parse non-foundational mathematics.
Apart from the foundational issues, the approach also has a problem with adaptivity of mathematical texts.

\subsection{Interchange Languages}

These approaches apply the general principles of formal languages while avoiding a commitment to a particular logic or implementation.
A major goal is system interoperability.

\textbf{Standardized representation languages} have been developed in the area of knowledge management such as \openmath \cite{openmath}, content \mathml \cite{mathml3}, and \omdoc \cite{omdoc}.
These prioritize standardizing the syntax using standard machine-friendly representation formats such as XML (where the formal structure of objects is explicit).
They do not specify user-friendly surface syntaxes (where the formal structure would have to be inferred through complex parsing and disambiguation) or rigorous semantics.
This allows their use as interchange languages (e.g., in the SCIEnce \cite{popcorn} and OpenDreamKit \cite{ODK:mitm:16} EU projects),
as a basis for integrating mathematics with the semantic web (e.g., in the MONET and HELM/MoWGLI FP6 projects), or as markup languages for web browsers (e.g., by the integration of MathML into HTML5).

A different trade-off is made in \textbf{interchange languages} mostly developed for theorem provers.
They are restricted to small families of logical languages used in theorem provers.
Thus, they are more widely applicable than individual logical languages but less widely than the truly universal standard representation languages.
The TPTP \cite{tptp} family of languages has played a major role in the community: it serves the role of a common language for automated theorem prover inputs and outputs.
TPTP was originally restricted to first-order logics, and a few extensions exist \cite{BRS:tptphol:08,tptp_tff_arithmetic}, which co-evolve with the available theorem provers, thus offering the possibility of problem exchange also between formal proof systems.
OpenTheory \cite{opentheory} is restricted to the HOL-based proof assistants.
It offers some support for abstracting from the systems' idiosyncrasies in order to increase portability, and some HOL theories have been manually refactored to make use of this abstraction.
The ISO-standardized Common Logic \cite{commonlogic} has a broader ambition, aiming at interchanging between any knowledge-based systems.
But its applicability to mathematics is limited by its focus on first-order logic and a lack of integration with mathematical software.

Overall, interchange languages focus mostly on a universal formal \emph{syntax} while sacrificing a universal semantics or restrict attention to small families of languages.
Neither provides strong support for type/module/proof/computation systems that would be critical to capture the complexity of large scale formal libraries.
A partial exception is the second author's \mmt system, which combines aspects of standard languages \cite{ODK:mitm:16,ODK:interop:17} and prover interchange languages \cite{BRS:tptphol:08,HR:tptp:12,KRS:thf1:16} with hard type and module systems \cite{RK:mmt:10}.
The OAF project \cite{KR:qed:14,KR:oafexp:20} used \mmt to represent large libraries of proof assistants in a standard representation language, including those of Mizar in \cite{IKR:mizar:11}, HOL Light in \cite{KR:hollight:14}, PVS in \cite{KMOR:pvs:17} (including the NASA library), Coq in \cite{MRS:coq:19} (including all available libraries), and Isabelle in \cite{KRW:isabelle:19} (including the Archive of Formal Proofs).

\section{Language Frameworks}\label{sec:languageframeworks}

Language frameworks are formal languages in which the syntax and semantics of other languages can be represented.
They are superficially related to parser frameworks but much stronger because they (i) allow specifying not only the syntax but also the semantics of a language, (ii) often offer strong support for context-sensitivity, which is critical in mathematics.

Logical frameworks are language frameworks for building formal language.
Examples are Isabelle \cite{isabelle}, Dedukti \cite{dedukti}, $\lambda$Prolog \cite{lambdaprolog}, or the LF \cite{lf} family including Twelf \cite{twelf} and others.
Frameworks also exist for building controlled natural languages such as GF \cite{gf}.

Contrary to the approaches discussed above, these frameworks do not in themselves provide languages for formalizing mathematics.
But they are worth discussing in this context for two reasons:
Firstly, they allow the rapid prototyping of implementations, which speeds up the feedback loop between language design and applications.
Thus, users can experiment with new languages and conduct large case studies in parallel.
Secondly, they allow developing scalable applications language-independently such that they are immediately applicable for any language defined in the framework.
That is important because evaluating formal languages often requires building (or trying to build) \emph{large} libraries in them.
Such applications include at least parsing and type-checking but can also include meta-reasoning (e.g., Twelf), interactive theorem proving (e.g., Isabelle), or language translation (e.g., GF).

Despite many successes in representing logical languages in logical frameworks (e.g., \cite{CHKMR:latinabs:11,KR:hollight:14,KMOR:pvs:17,MRS:coq:19}), current frameworks cover only unrealistically simple languages compared to the needs for mathematically structured content and do not have good support for, e.g., soft type systems and soft module systems and practical proof systems.
Thus, even the representation of the already insufficient languages discussed above is often very difficult or not possible.

Therefore, more flexible logical frameworks were developed recently.
Both ELPI \cite{elpi} and \mmt \cite{rabe:howto:14,rabe:recon:17} allow users to flexibly change critical algorithms whenever a concrete language definition needs it.
That makes them more promising for representing languages designed for mathematical content (and can even allow sharing some functionality across incompatible foundations).

\mmt~\cite{rabe:combining:10,RK:mmt:10} is a logic-independent representation and management system for formal logical content that uses logical frameworks to provide a rigorous semantics for \omdoc and \openmath.
It manages all aspects of language design in a coherent framework including language definition, rapid prototyping of tools, and library development. Fully parametric in the choice of formal system, it maximizes the reuse of concepts, formalizations, and tool support.
It subsumes in particular logical frameworks such as the LF family \cite{MR:prototyping:19}.
The LATIN project \cite{CHKMR:latinabs:11} used an \mmt precursor language based on the Twelf module system \cite{RS:twelfmod:09} to build a library of common logics of symbolic software systems and proof checkers. It contains close to 1000 modules (theories and morphisms between them), which can be imported into \mmt.

\cite{KRSS:mmtelpi:20} makes the first steps towards combining the advantages of \mmt and ELPI.
\cite{glf} uses \mmt to extend LF-like logical frameworks with the natural language framework GF.

\section{Interchange Libraries}\label{sec:lib}

The quest for the best formal language for mathematics is likely to never-ending.
Therefore, it is important to investigate how to combine the existing libraries of formalized content.
Due to major incompatibilities between the various formal systems, this is an extremely difficult problem, and it would go beyond the scope of this paper to discuss approaches in detail.
But we want to mention the idea of interchange libraries because we consider it to be one of the most promising ideas.

An interchange library $I$ is a formalization of mathematics written in an intermediate language with the goal of serving as an interoperability layer between formal systems.
The main idea is that all translations from source system $S$ to target system $T$ are split into two steps $S\to I$ and $I\to T$.

Both steps have characteristic difficulties.
The step $S\to I$ is usually a partial translation because every formal systems uses idiosyncratic features that cannot be represented in $I$ and optimizations for verification/computation that need not be represented in $I$.
The step $I\to T$ tends to be easier, but there is a tricky trade-off in the design of $I$: the less $I$ commits to a particular formal system, the more systems $T$ can be handled but the more difficult the individual translations $I\to T$ become.
In practice, a further major logistic problem is that $I$ and the translations via it needs to be built and maintained, which is even harder to organize and fund than for the systems $S$ and $T$ themselves.

The standard content dictionaries written in OpenMath \cite{openmath} were the first concerted effort to build an interchange library.
$214$ dictionaries (including contributed ones) declaring $1578$ symbols are maintained by the OpenMath Society.
These focus on declaring names for mathematical symbols and describing their semantics verbally and with formal axioms.
However, the approach was not widely adopted as little tool support existed for OpenMath itself and for OpenMath-based interoperability.
Individual formal systems were also less able to export/import their objects at all.

Recently, the idea was picked up again in the OpenDreamKit project.
It uses \mmt (whose language of theories and expressions essentially subsumes OpenMath CDs and objects) to write a formal interchange library (dubbed MitM for Math-in-the-middle) \cite{ODK:mitm:16}.
MitM is more formal than the OpenMath CDs, in particular employing a hard type and module system.
It was used as an interoperability layer for computer algebra systems  \cite{ODK:interop:17} and mathematical databases \cite{WKR:virtual:17,BKR:data:19}.

A complementary approach is SMGloM \cite{smglom}, a multi-lingual glossary of mathematical concepts.
It retains the untyped natural of OpenMath CDs but uses sTeX to obtain tool support for writing the library.

SMGloM and MitM serve similar purposes with different methods that recall the distinctions described in Section~\ref{sec:overview}:
SMGloM uses mostly natural language, and MitM uses formal language with hard type and module system.
The short-comings of these efforts seem to indicate that soft types and modules may be the best trade-off for building an interchange library.

In order to streamline the process of building the translations $S\to I$ and $I\to T$, the concept of \emph{alignments} was developed \cite{KKMR:alignments:16}.
An alignment between two symbols $c$ and $c'$ in different libraries captures that translations should try to translate objects with $c$ to objects with head $d$.
Both exact manual efforts \cite{MRLR:alignments:17} and machine learning--based heuristic approaches were used to find alignments across formal libraries.
The latter includes alignment from six proof assistants~\cite{tgck-jsc19}, showing that such alignments allow both conjecturing and more powerful automation~\cite{tgck-lpar15}.
The same approach has been used to obtain alignments between informal and formal libraries, which can be used to automatically
formalize parts of mathematical texts, both statistically~\cite{ckjujv-itp17} and using deep learning techniques~\cite{qwckju-cicm18}.
Similarly, \cite{nnexus} automatically obtains alignments between informal libraries.

\section{Conclusion}\label{sec:conc}
We have presented a survey of languages for formalizing mathematics.
The various languages have been designed and implemented for different purposes and have different features, and their many distinguishing features give them characteristic advantages and disadvantages.
Natural language that mathematicians are used to lacks formal semantics (and in many cases even formal syntax).
But fully formal languages are still very far from natural language.
And existing intermediate languages lack complex structuring features and large libraries and scalable tools that would make them directly usable for formalization.

We expect that future research in the domain must continue to experiment with language development aiming at the formal representation of syntax and semantics while preserving natural readability and extensibility and large-scale structuring features.
The use of language frameworks will be helpful to rapidly experiment with these novel ideas.
We see a lot of potential in the development of a new intermediate language along those lines that could enable partial and stepwise formalization as well as provide an interoperability layer for formal languages.
Concretely, we expect this future language to feature at least a combination of soft type and module systems with rigorous development of their hard analogues as emergent features.

\newcommand{\etalchar}[1]{$^{#1}$}


\begin{thebibliography}{GGMR09}

\bibitem[ABC{\etalchar{+}}10]{mathml3}
R.~Ausbrooks, S.~Buswell, D.~Carlisle, G.~Chavchanidze, S.~Dalmas, S.~Devitt,
  A.~Diaz, S.~Dooley, R.~Hunter, P.~Ion, M.~Kohlhase, A.~Lazrek, P.~Libbrecht,
  B.~Miller, R.~Miner, C.~Rowley, M.~Sargent, B.~Smith, N.~Soiffer, R.~Sutor,
  and S.~Watt.
\newblock {Mathematical Markup Language (MathML) Version 3.0}.
\newblock Technical report, World Wide Web Consortium, 2010.
\newblock See \url{http://www.w3.org/TR/MathML3}.

\bibitem[Abr96]{bmethod}
J.~Abrial.
\newblock {\em {The B-Book: Assigning Programs to Meanings}}.
\newblock Cambridge University Press, 1996.

\bibitem[Art]{proofpower}
R.~Arthan.
\newblock {ProofPower}.
\newblock \url{http://www.lemma-one.com/ProofPower/}.

\bibitem[ASTZ06]{matita}
A.~Asperti, C.~{Sacerdoti Coen}, E.~Tassi, and S.~Zacchiroli.
\newblock {Crafting a Proof Assistant}.
\newblock In T.~Altenkirch and C.~McBride, editors, {\em TYPES}, pages 18--32.
  Springer, 2006.

\bibitem[BCC{\etalchar{+}}04]{openmath}
S.~Buswell, O.~Caprotti, D.~Carlisle, M.~Dewar, M.~Gaetano, and M.~Kohlhase.
\newblock {The Open Math Standard, Version 2.0}.
\newblock Technical report, The Open Math Society, 2004.
\newblock See \url{http://www.openmath.org/standard/om20}.

\bibitem[BCH12]{dedukti}
M.~Boespflug, Q.~Carbonneaux, and O.~Hermant.
\newblock The {$\lambda\Pi$}-calculus modulo as a universal proof language.
\newblock In D.~Pichardie and T.~Weber, editors, {\em Proceedings of PxTP2012:
  Proof Exchange for Theorem Proving}, pages 28--43, 2012.

\bibitem[BKP19]{cbckkp-itp19}
C.~Brown, C.~Kaliszyk, and K.~P\k{a}k.
\newblock Higher-order {T}arski {G}rothendieck as a foundation for formal
  proof.
\newblock In John Harrison, John O'Leary, and Andrew Tolmach, editors, {\em
  Interactive Theorem Proving}, volume 141 of {\em LIPIcs}, pages 9:1--9:16,
  2019.

\bibitem[BKR19]{BKR:data:19}
K.~Ber{\v{c}}i{\v{c}}, M.~Kohlhase, and F.~Rabe.
\newblock {Towards a Unified Mathematical Data Infrastructure: Database and
  Interface Generation}.
\newblock In C.~Kaliszyk, E.~Brady, A.~Kohlhase, and C.~{Sacerdoti Coen},
  editors, {\em Intelligent Computer Mathematics}, pages 28--43. Springer,
  2019.

\bibitem[BRS08]{BRS:tptphol:08}
C.~Benzm{\"u}ller, F.~Rabe, and G.~Sutcliffe.
\newblock {THF0 -- The core of the TPTP Language for Higher-Order Logic}.
\newblock In A.~Armando, P.~Baumgartner, and G.~Dowek, editors, {\em 4th
  International Joint Conference on Automated Reasoning}, pages 491--506.
  Springer, 2008.

\bibitem[CAB{\etalchar{+}}86]{nuprl}
R.~Constable, S.~Allen, H.~Bromley, W.~Cleaveland, J.~Cremer, R.~Harper,
  D.~Howe, T.~Knoblock, N.~Mendler, P.~Panangaden, J.~Sasaki, and S.~Smith.
\newblock {\em {Implementing Mathematics with the Nuprl Development System}}.
\newblock Prentice-Hall, 1986.

\bibitem[CF58]{curry}
H.~Curry and R.~Feys.
\newblock {\em Combinatory Logic}.
\newblock North-Holland, Amsterdam, 1958.

\bibitem[CFK{\etalchar{+}}09]{naproche}
M.~Cramer, B.~Fisseni, P.~Koepke, D.~K{\"u}hlwein, B.~Schr{\"o}der, and
  J.~Veldman.
\newblock {The Naproche Project Controlled Natural Language Proof Checking of
  Mathematical Texts}.
\newblock In N.~Fuchs, editor, {\em Controlled Natural Language}, pages
  170--186. Springer, 2009.

\bibitem[CH88]{calcconstructions}
T.~Coquand and G.~Huet.
\newblock {The Calculus of Constructions}.
\newblock {\em {Information and Computation}}, 76(2/3):95--120, 1988.

\bibitem[CHK{\etalchar{+}}11]{CHKMR:latinabs:11}
M.~Codescu, F.~Horozal, M.~Kohlhase, T.~Mossakowski, and F.~Rabe.
\newblock {Project Abstract: Logic Atlas and Integrator (LATIN)}.
\newblock In J.~Davenport, W.~Farmer, F.~Rabe, and J.~Urban, editors, {\em
  Intelligent Computer Mathematics}, pages 289--291. Springer, 2011.

\bibitem[Chu40]{churchtypes}
A.~Church.
\newblock {A Formulation of the Simple Theory of Types}.
\newblock {\em {Journal of Symbolic Logic}}, 5(1):56--68, 1940.

\bibitem[{Coq}15]{coq}
{Coq Development Team}.
\newblock {The Coq Proof Assistant: Reference Manual}.
\newblock Technical report, INRIA, 2015.

\bibitem[DGST15]{elpi}
C.~Dunchev, F.~Guidi, C.~{Sacerdoti Coen}, and E.~Tassi.
\newblock {ELPI: Fast, Embeddable, lambda-Prolog Interpreter}.
\newblock In M.~Davis, A.~Fehnker, A.~McIver, and A.~Voronkov, editors, {\em
  Logic for Programming, Artificial Intelligence, and Reasoning}, pages
  460--468, 2015.

\bibitem[DIK{\etalchar{+}}16]{ODK:mitm:16}
P.~Dehaye, M.~Iancu, M.~Kohlhase, A.~Konovalov, S.~Leli{\`e}vre, D.~M{\"u}ller,
  M.~Pfeiffer, F.~Rabe, N.~Thi{\'e}ry, and T.~Wiesing.
\newblock {Interoperability in the OpenDreamKit Project: The Math-in-the-Middle
  Approach}.
\newblock In M.~Kohlhase, L.~{de Moura}, M.~Johansson, B.~Miller, and F.~Tompa,
  editors, {\em Intelligent Computer Mathematics}, pages 117--131. Springer,
  2016.

\bibitem[dKA{\etalchar{+}}15]{lean}
L.~{de Moura}, S.~Kong, J.~Avigad, F.~{van Doorn}, and J.~{von Raumer}.
\newblock {The Lean Theorem Prover (System Description)}.
\newblock In A.~Felty and A.~Middeldorp, editors, {\em Automated Deduction},
  pages 378--388. Springer, 2015.

\bibitem[edi07]{commonlogic}
Common~Logic editors.
\newblock {Common Logic (CL) — A framework for a family of logic-based
  languages}.
\newblock Technical Report 24707, ISO/IEC, 2007.

\bibitem[FGT93]{imps}
W.~Farmer, J.~Guttman, and F.~Thayer.
\newblock {IMPS: An Interactive Mathematical Proof System}.
\newblock {\em {Journal of Automated Reasoning}}, 11(2):213--248, 1993.

\bibitem[GAA{\etalchar{+}}13]{DBLP:conf/itp/GonthierAABCGRMOBPRSTT13}
G.~Gonthier, A.~Asperti, J.~Avigad, Y.~Bertot, C.~Cohen, F.~Garillot, S.~Le
  Roux, A.~Mahboubi, R.~O'Connor, S.~Ould Biha, I.~Pasca, L.~Rideau,
  A.~Solovyev, E.~Tassi, and L.~Th{\'e}ry.
\newblock A machine-checked proof of the {Odd Order Theorem}.
\newblock In S.~Blazy, C.~Paulin-Mohring, and D.~Pichardie, editors, {\em
  Interactive Theorem Proving (ITP)}, volume 7998 of {\em LNCS}, pages
  163--179. Springer, 2013.

\bibitem[Gan13]{Ganesalingam}
M.~Ganesalingam.
\newblock {\em The Language of Mathematics. A Linguistic and Philosophical
  Investigation}.
\newblock Number 7805 in LNCS. Springer, 2013.

\bibitem[GC14]{nnexus}
D.~Ginev and J.~Corneli.
\newblock {NNexus} reloaded.
\newblock In S.~Watt, J.~Davenport, A.~Sexton, P.~Sojka, and J.~Urban, editors,
  {\em Intelligent Computer Mathematics}, pages 423--426. Springer, 2014.

\bibitem[GGMR09]{gonthier_packaging}
F.~Garillot, G.~Gonthier, A.~Mahboubi, and L.~Rideau.
\newblock Packaging mathematical structures.
\newblock In S.~Berghofer, T.~Nipkow, C.~Urban, and M.~Wenzel, editors, {\em
  Theorem Proving in Higher Order Logics}, pages 327--342. Springer, 2009.

\bibitem[GIJ{\etalchar{+}}16]{smglom}
D.~Ginev, M.~Iancu, C.~Jucovschi, A.~Kohlhase, M.~Kohlhase, A.~Oripov,
  J.~Schefter, W.~Sperber, O.~Teschke, and T.~Wiesing.
\newblock The {SMGloM} project and system: Towards a terminology and ontology
  for mathematics.
\newblock In G.{-}M. Greuel, T.~Koch, P.~Paule, and A.J. Sommese, editors, {\em
  Mathematical Software - {ICMS} 2016 - 5th International Conference, Berlin,
  Germany, July 11-14, 2016, Proceedings}, volume 9725 of {\em LNCS}, pages
  451--457. Springer, 2016.

\bibitem[GK15]{tgck-lpar15}
T.~Gauthier and C.~Kaliszyk.
\newblock Sharing {HOL4} and {HOL L}ight proof knowledge.
\newblock In M.~Davis, A.~Fehnker, A.~McIver, and A.~Voronkov, editors, {\em
  20th International Conference on Logic for Programming, Artificial
  Intelligence, and Reasoning (LPAR 2015)}, volume 9450 of {\em LNCS}, pages
  372--386. Springer, 2015.

\bibitem[GK19]{tgck-jsc19}
T.~Gauthier and C.~Kaliszyk.
\newblock Aligning concepts across proof assistant libraries.
\newblock {\em J. Symbolic Computation}, 90:89--123, 2019.

\bibitem[Hal12]{halesdense}
T.~Hales.
\newblock {\em Dense Sphere Packings: A Blueprint for Formal Proofs}, volume
  400 of {\em London Mathematical Society Lecture Note Series}.
\newblock Cambridge University Press, 2012.

\bibitem[Hal14]{Hales-Developments}
T.~Hales.
\newblock Developments in formal proofs.
\newblock {\em Séminaire Bourbaki}, 1086, 2013--2014.
\newblock \href{http://arxiv.org/abs/1408.6474}{abs/1408.6474}.

\bibitem[Hal17]{fabstracts}
T.~Hales.
\newblock The formal abstracts project, 2017.
\newblock \url{https://formalabstracts.github.io/}.

\bibitem[Har96]{hollight}
J.~Harrison.
\newblock {HOL Light: A Tutorial Introduction}.
\newblock In {\em {Proceedings of the First International Conference on Formal
  Methods in Computer-Aided Design}}, pages 265--269. Springer, 1996.

\bibitem[HHP93]{lf}
R.~Harper, F.~Honsell, and G.~Plotkin.
\newblock {A framework for defining logics}.
\newblock {\em {Journal of the Association for Computing Machinery}},
  40(1):143--184, 1993.

\bibitem[{HOL}]{hol4}
{HOL4 development team}.
\newblock \url{https://hol-theorem-prover.org/}.

\bibitem[How80]{howard}
W.~Howard.
\newblock The formulas-as-types notion of construction.
\newblock In {\em To H.B. Curry: Essays on Combinatory Logic, Lambda-Calculus
  and Formalism}, pages 479--490. Academic Press, 1980.

\bibitem[HR09]{popcorn}
P.~Horn and D.~Roozemond.
\newblock {OpenMath in SCIEnce: SCSCP and POPCORN}.
\newblock In J.~Carette, L.~Dixon, C.~{Sacerdoti Coen}, and S.~Watt, editors,
  {\em Intelligent Computer Mathematics}, pages 474--479. Springer, 2009.

\bibitem[HR15]{HR:tptp:12}
F.~Horozal and F.~Rabe.
\newblock {Formal Logic Definitions for Interchange Languages}.
\newblock In M.~Kerber, J.~Carette, C.~Kaliszyk, F.~Rabe, and V.~Sorge,
  editors, {\em Intelligent Computer Mathematics}, pages 171--186. Springer,
  2015.

\bibitem[Hum12]{mathnat}
M.~Humayoun.
\newblock {\em {Developing System MathNat for Automatic Formalization of
  Mathematical texts}}.
\newblock PhD thesis, {Universit{\'e} de Grenoble}, 2012.

\bibitem[Hur09]{opentheory}
J.~Hurd.
\newblock {OpenTheory: Package Management for Higher Order Logic Theories}.
\newblock In G.~Dos Reis and L.~Th{\'e}ry, editors, {\em {Programming Languages
  for Mechanized Mathematics Systems}}, pages 31--37. {ACM}, 2009.

\bibitem[IKR11]{IKR:mizar:11}
M.~Iancu, M.~Kohlhase, and F.~Rabe.
\newblock {Translating the Mizar Mathematical Library into OMDoc format}.
\newblock Technical Report KWARC Report-01/11, Jacobs University Bremen, 2011.

\bibitem[KAE{\etalchar{+}}14]{SeL414}
G.~Klein, J.~Andronick, K.~Elphinstone, T.C. Murray, T.~Sewell, R.~Kolanski,
  and G.~Heiser.
\newblock Comprehensive formal verification of an {OS} microkernel.
\newblock {\em ACM Trans. Comput. Syst.}, 32(1):2, 2014.

\bibitem[KKMR16]{KKMR:alignments:16}
C.~Kaliszyk, M.~Kohlhase, D.~M{\"u}ller, and F.~Rabe.
\newblock {A Standard for Aligning Mathematical Concepts}.
\newblock In A.~Kohlhase, M.~Kohlhase, P.~Libbrecht, B.~Miller, F.~Tompa,
  A.~Naummowicz, W.~Neuper, P.~Quaresma, and M.~Suda, editors, {\em Work in
  Progress at CICM 2016}, pages 229--244. CEUR-WS.org, 2016.

\bibitem[KMM00]{acl2}
M.~Kaufmann, P.~Manolios, and J~Moore.
\newblock {\em {Computer-Aided Reasoning: An Approach}}.
\newblock Kluwer Academic Publishers, 2000.

\bibitem[KMOR17]{KMOR:pvs:17}
M.~Kohlhase, D.~M{\"u}ller, S.~Owre, and F.~Rabe.
\newblock {Making PVS Accessible to Generic Services by Interpretation in a
  Universal Format}.
\newblock In M.~Ayala-Rincon and C.~Munoz, editors, {\em Interactive Theorem
  Proving}, pages 319--335. Springer, 2017.

\bibitem[KMP{\etalchar{+}}17]{ODK:interop:17}
M.~Kohlhase, D.~M{\"u}ller, M.~Pfeiffer, F.~Rabe, N.~Thi{\'e}ry, V.~Vasilyev,
  and T.~Wiesing.
\newblock {Knowledge-Based Interoperability for Mathematical Software Systems}.
\newblock In J.~Bl{\"o}mer, I.~Kotsireas, T.~Kutsia, and D.~Simos, editors,
  {\em Mathematical Aspects of Computer and Information Sciences}, pages
  195--210. Springer, 2017.

\bibitem[KN12]{kofler_parser}
K.~Kofler and A.~Neumaier.
\newblock Dyngenpar - a dynamic generalized parser for common mathematical
  language.
\newblock In J.~Jeuring, J.~Campbell, J.~Carette, G.~Dos Reis, P.~Sojka,
  M.~Wenzel, and V.~Sorge, editors, {\em Intelligent Computer Mathematics},
  volume 7362, pages 386--401. Springer, 2012.

\bibitem[Knu84]{literate_programming}
D.~Knuth.
\newblock Literate programming.
\newblock {\em The Computer Journal}, 27(2):97--111, 1984.

\bibitem[Koh06]{omdoc}
M.~Kohlhase.
\newblock {\em {OMDoc: An Open Markup Format for Mathematical Documents
  (Version 1.2)}}.
\newblock Number 4180 in {Lecture Notes in Artificial Intelligence}. Springer,
  2006.

\bibitem[Koh08]{stex}
M.~Kohlhase.
\newblock {Using {\LaTeX} as a Semantic Markup Format}.
\newblock {\em Mathematics in Computer Science}, 2(2):279--304, 2008.

\bibitem[KP19a]{ckkp-jar19}
C.~Kaliszyk and K.~P\k{a}k.
\newblock Semantics of {M}izar as an {I}sabelle object logic.
\newblock {\em J. Autom. Reasoning}, 63(3):557--595, 2019.

\bibitem[KP19b]{ckkp-itp19}
C.~Kaliszyk and K.~Pąk.
\newblock Declarative proof translation (short paper).
\newblock In J.~Harrison, J.~O'Leary, and A.~Tolmach, editors, {\em 10th
  International Conference on Interactive Theorem Proving (ITP 2019)}, volume
  141 of {\em LIPIcs}, pages 35:1--35:7, 2019.

\bibitem[KR93]{KampReyle}
H.~Kamp and U.~Reyle.
\newblock {\em From Discourse to Logic. Introduction to Modeltheoretic
  Semantics of Natural Language, Formal Logic and Discourse Representation
  Theory}, volume~42 of {\em Studies in Linguistics and Philosophy}.
\newblock Springer, 1993.

\bibitem[KR14]{KR:hollight:14}
C.~Kaliszyk and F.~Rabe.
\newblock {Towards Knowledge Management for HOL Light}.
\newblock In S.~Watt, J.~Davenport, A.~Sexton, P.~Sojka, and J.~Urban, editors,
  {\em Intelligent Computer Mathematics}, pages 357--372. Springer, 2014.

\bibitem[KR16]{KR:qed:14}
M.~Kohlhase and F.~Rabe.
\newblock {QED Reloaded: Towards a Pluralistic Formal Library of Mathematical
  Knowledge}.
\newblock {\em Journal of Formalized Reasoning}, 9(1):201--234, 2016.

\bibitem[KR20]{KR:oafexp:20}
M.~Kohlhase and F.~Rabe.
\newblock {Experiences from Exporting Major Proof Assistant Libraries}.
\newblock see \url{https://kwarc.info/people/frabe/Research/KR_oafexp_20.pdf},
  2020.

\bibitem[KRS16]{KRS:thf1:16}
C.~Kaliszyk, F.~Rabe, and G.~Sutcliffe.
\newblock {TH1: The TPTP Typed Higher-Order Form with Rank-1 Polymorphism}.
\newblock In P.~Fontaine, S.~Schulz, and J.~Urban, editors, {\em Workshop on
  Practical Aspects of Automated Reasoning}, pages 41--55, 2016.

\bibitem[KRSS20]{KRSS:mmtelpi:20}
M.~Kohlhase, F.~Rabe, C.~{Sacerdoti Coen}, and J.~Schaefer.
\newblock {Logic-Independent Proof Search in Logical Frameworks (short paper)}.
\newblock In N.~Peltier and V.~Sofronie-Stokkermans, editors, {\em Automated
  Reasoning}. Springer, 2020.

\bibitem[KRW20]{KRW:isabelle:19}
M.~Kohlhase, F.~Rabe, and M.~Wenzel.
\newblock {Making Isabelle Content Accessible in Knowledge in Representation
  Formats}.
\newblock see
  \url{https://kwarc.info/people/frabe/Research/KRW_isabelle_19.pdf}, 2020.

\bibitem[KS19]{glf}
M.~Kohlhase and J.~Schaefer.
\newblock {GF + MMT = GLF -- From Language to Semantics through LF}.
\newblock In D.~Miller and I.~Scagnetto, editors, {\em Logical Frameworks and
  Meta-Languages: Theory and Practice}, pages 24--39. Open Publishing
  Association, 2019.

\bibitem[KUV17]{ckjujv-itp17}
C.~Kaliszyk, J.~Urban, and J.~Vysko\v{c}il.
\newblock Automating formalization by statistical and semantic parsing of
  mathematics.
\newblock In M.~Ayala{-}Rinc{\'{o}}n and C.A. Mu{\~{n}}oz, editors, {\em
  Interactive Theorem Proving}, volume 10499 of {\em LNCS}, pages 12--27.
  Springer, 2017.

\bibitem[KW08]{mathlang}
F.~Kamareddine and J.~Wells.
\newblock Computerizing mathematical text with mathlang.
\newblock In M.~Ayala-Rinc{\'o}n and E.~Haeusler, editors, {\em Logical and
  Semantic Frameworks, with Applications}, pages 5--30. ENTCS, 2008.

\bibitem[KWZB14]{mathlang_handbook}
F.~Kamareddine, J.~Wells, C.~Zengler, and H.~Barendregt.
\newblock Computerising mathematical text.
\newblock In J.~Siekmann, editor, {\em Computational Logic}. Elsevier, 2014.

\bibitem[Ler09]{Compcert09}
X.~Leroy.
\newblock Formal verification of a realistic compiler.
\newblock {\em Commun. ACM}, 52(7):107--115, 2009.

\bibitem[Lin07]{gap}
S.~Linton.
\newblock {GAP}: groups, algorithms, programming.
\newblock {\em ACM Communications in Computer Algebra}, 41(3):108--109, 2007.

\bibitem[Meg07]{metamath}
N.~Megill.
\newblock {\em Metamath: A Computer Language for Pure Mathematics}.
\newblock Lulu Press, Morrisville, North Carolina, 2007.

\bibitem[Mil72]{lcf}
R.~Milner.
\newblock Logic for computable functions: descriptions of a machine
  implementation.
\newblock {\em ACM SIGPLAN Notices}, 7:1--6, 1972.

\bibitem[ML74]{martinlof}
P.~Martin-{L\"o}f.
\newblock {An Intuitionistic Theory of Types: Predicative Part}.
\newblock In {\em {Proceedings of the '73 Logic Colloquium}}, pages 73--118.
  North-Holland, 1974.

\bibitem[MN86]{lambdaprolog}
D.~Miller and G.~Nadathur.
\newblock Higher-order logic programming.
\newblock In E.~Shapiro, editor, {\em Proceedings of the Third International
  Conference on Logic Programming}, pages 448--462. Springer, 1986.

\bibitem[MR19]{MR:prototyping:19}
D.~M{\"u}ller and F.~Rabe.
\newblock {Rapid Prototyping Formal Systems in MMT: Case Studies}.
\newblock In D.~Miller and I.~Scagnetto, editors, {\em Logical Frameworks and
  Meta-languages: Theory and Practice}, pages 40--54, 2019.

\bibitem[MRK18]{MRK:modelsof:18}
D.~M{\"u}ller, F.~Rabe, and M.~Kohlhase.
\newblock {Theories as Types}.
\newblock In D.~Galmiche, S.~Schulz, and R.~Sebastiani, editors, {\em Automated
  Reasoning}, pages 575--590. Springer, 2018.

\bibitem[MRLR17]{MRLR:alignments:17}
D.~M{\"u}ller, C.~Rothgang, Y.~Liu, and F.~Rabe.
\newblock {Alignment-based Translations Across Formal Systems Using Interface
  Theories}.
\newblock In C.~Dubois and B.~{Woltzenlogel Paleo}, editors, {\em Proof
  eXchange for Theorem Proving}, pages 77--93. Open Publishing Association,
  2017.

\bibitem[MRS19]{MRS:coq:19}
D.~M{\"u}ller, F.~Rabe, and C.~{Sacerdoti Coen}.
\newblock {The Coq Library as a Theory Graph}.
\newblock In C.~Kaliszyk, E.~Brady, A.~Kohlhase, and C.~{Sacerdoti Coen},
  editors, {\em Intelligent Computer Mathematics}, pages 171--186. Springer,
  2019.

\bibitem[Nau14]{Naumowicz14}
Adam Naumowicz.
\newblock {SAT}-enhanced {M}izar proof checking.
\newblock In Stephen~M. Watt, James~H. Davenport, Alan~P. Sexton, Petr Sojka,
  and Josef Urban, editors, {\em Intelligent Computer Mathematics -
  International Conference, {CICM} 2014, Coimbra, Portugal, July 7-11, 2014.
  Proceedings}, volume 8543 of {\em LNCS}, pages 449--452. Springer, 2014.

\bibitem[Nor05]{agda}
U.~Norell.
\newblock {The Agda WiKi}, 2005.
\newblock \url{http://wiki.portal.chalmers.se/agda}.

\bibitem[NPW02]{isabellehol}
T.~Nipkow, L.~Paulson, and M.~Wenzel.
\newblock {\em {Isabelle/HOL --- A Proof Assistant for Higher-Order Logic}}.
\newblock Springer, 2002.

\bibitem[ORS92]{pvs}
S.~Owre, J.~Rushby, and N.~Shankar.
\newblock {PVS: A Prototype Verification System}.
\newblock In D.~Kapur, editor, {\em {11th International Conference on Automated
  Deduction (CADE)}}, pages 748--752. Springer, 1992.

\bibitem[Pau94]{isabelle}
L.~Paulson.
\newblock {\em {Isabelle: A Generic Theorem Prover}}, volume 828 of {\em
  Lecture Notes in Computer Science}.
\newblock Springer, 1994.

\bibitem[PC93]{isabelle_zf}
L.~Paulson and M.~Coen.
\newblock {Zermelo-Fraenkel Set Theory}, 1993.
\newblock {Isabelle distribution, ZF/ZF.thy}.

\bibitem[PS99]{twelf}
F.~Pfenning and C.~Sch{\"u}rmann.
\newblock System description: {Twelf} - a meta-logical framework for deductive
  systems.
\newblock In H.~Ganzinger, editor, {\em Automated Deduction}, pages 202--206,
  1999.

\bibitem[Rab13]{rabe:combining:10}
F.~Rabe.
\newblock {A Logical Framework Combining Model and Proof Theory}.
\newblock {\em Mathematical Structures in Computer Science}, 23(5):945--1001,
  2013.

\bibitem[Rab17]{rabe:howto:14}
F.~Rabe.
\newblock {How to Identify, Translate, and Combine Logics?}
\newblock {\em Journal of Logic and Computation}, 27(6):1753--1798, 2017.

\bibitem[Rab18]{rabe:recon:17}
F.~Rabe.
\newblock {A Modular Type Reconstruction Algorithm}.
\newblock {\em ACM Transactions on Computational Logic}, 19(4):1--43, 2018.

\bibitem[Ran11]{gf}
A.~Ranta.
\newblock {\em {Grammatical Framework: Programming with Multilingual
  Grammars}}.
\newblock {CSLI Publications}, 2011.

\bibitem[RK13]{RK:mmt:10}
F.~Rabe and M.~Kohlhase.
\newblock {A Scalable Module System}.
\newblock {\em Information and Computation}, 230(1):1--54, 2013.

\bibitem[RS09]{RS:twelfmod:09}
F.~Rabe and C.~Sch{\"u}rmann.
\newblock {A Practical Module System for LF}.
\newblock In J.~Cheney and A.~Felty, editors, {\em Proceedings of the Workshop
  on Logical Frameworks: Meta-Theory and Practice (LFMTP)}, pages 40--48. ACM
  Press, 2009.

\bibitem[S{\etalchar{+}}13]{sage}
W.~Stein et~al.
\newblock {\em {Sage Mathematics Software}}.
\newblock The Sage Development Team, 2013.
\newblock \url{http://www.sagemath.org}.

\bibitem[SSCB12]{tptp_tff_arithmetic}
G.~Sutcliffe, S.~Schulz, K.~Claessen, and P.~Baumgartner.
\newblock {The TPTP Typed First-Order Form with Arithmetic}.
\newblock In N.~Bj{\o}rner and A.~Voronkov, editors, {\em Logic for
  Programming, Artificial Intelligence}, pages 406--419. Springer, 2012.

\bibitem[Sut09]{tptp}
G.~Sutcliffe.
\newblock {The TPTP Problem Library and Associated Infrastructure: The FOF and
  CNF Parts, v3.5.0}.
\newblock {\em Journal of Automated Reasoning}, 43(4):337--362, 2009.

\bibitem[TB85]{mizar}
A.~Trybulec and H.~Blair.
\newblock {Computer Assisted Reasoning with MIZAR}.
\newblock In A.~Joshi, editor, {\em Proceedings of the 9th International Joint
  Conference on Artificial Intelligence}, pages 26--28. Morgan Kaufmann, 1985.

\bibitem[Wen11]{isabelle_documentoriented}
M.~Wenzel.
\newblock Isabelle as a document-oriented proof assistant.
\newblock In J.~Davenport, W.~Farmer, J.~Urban, and F.~Rabe, editors, {\em
  Intelligent Computer Mathematics}, pages 244--259. Springer, 2011.

\bibitem[Wie07]{qedrevisited}
F.~Wiedijk.
\newblock {The QED Manifesto Revisited}.
\newblock In {\em {From Insight to Proof, Festschrift in Honour of Andrzej
  Trybulec}}, pages 121--133, 2007.

\bibitem[Win14]{theorema}
W.~Windsteiger.
\newblock {Theorema 2.0: A System for Mathematical Theory Exploration}.
\newblock In H.~Hong and C.~Yap, editors, {\em International Congress on
  Mathematical Software}, pages 49--52. Springer, 2014.

\bibitem[WKR17]{WKR:virtual:17}
T.~Wiesing, M.~Kohlhase, and F.~Rabe.
\newblock {Virtual Theories -- A Uniform Interface to Mathematical Knowledge
  Bases}.
\newblock In J.~Bl{\"o}mer, I.~Kotsireas, T.~Kutsia, and D.~Simos, editors,
  {\em Mathematical Aspects of Computer and Information Sciences}, pages
  243--257. Springer, 2017.

\bibitem[WKU18]{qwckju-cicm18}
Q.~Wang, C.~Kaliszyk, and J.~Urban.
\newblock First experiments with neural translation of informal to formal
  mathematics.
\newblock In F.~Rabe, W.M. Farmer, G.O. Passmore, and A.~Youssef, editors, {\em
  Intelligent Computer Mathematics}, volume 11006 of {\em LNCS}, pages
  255--270. Springer, 2018.

\bibitem[Wol12]{mathematica}
Wolfram.
\newblock Mathematica, 2012.

\end{thebibliography}
\end{document}